\documentclass[12pt,preprint]{aastex}

\slugcomment{}

\shorttitle{Cool Subdwarfs Investigations II: Multiplicity}
\shortauthors{Jao et al.}

\begin{document}

\title{Cool Subdwarf Investigations II: Multiplicity}

\author{Wei-Chun Jao}

\affil{Department of Physics and Astronomy, Georgia State University,
Atlanta, GA 30302-4106} \email{jao@chara.gsu.edu}

\author{Brian D. Mason\altaffilmark{1,2}, William I
Hartkopf\altaffilmark{1,2}}

\affil{US Naval Observatory, 3450 Massachusetts Avenue NW, Washington,
DC 20392-5420} \email{bdm, wih@usno.navy.mil}

\author{Todd J. Henry and Stephanie N. Ramos}

\affil{Department of Physics and Astronomy, Georgia State University,
Atlanta, GA 30302-4106} \email{thenry@chara.gsu.edu,
sramos1@student.gsu.edu}

\altaffiltext{1}{Visiting Astronomer, Cerro Tololo Inter-American
Observatory.  CTIO is operated by AURA, Inc.\ under contract to the
National Science Foundation.}

\altaffiltext{2}{Visiting Astronomer, Kitt Peak National Observatory.
KPNO is operated by AURA, Inc.\ under contract to the National Science
Foundation.}

\begin{abstract}

Cool subdwarfs of types K and M are the fainter counterparts of cool
main sequence dwarfs that dominate the Galactic population.  In this
paper we present the results of an optical speckle survey of 62
confirmed cool subdwarf systems within 60 pc.  We have resolved two
new companions and confirmed two previously known companions with
separations 0\farcs13 to 3\farcs29.  After including previously known
wide companions and all known spectroscopic binaries, we determine the
multiplicity rate of cool subdwarfs to be 26$\pm$6\%, which is
somewhat lower than comparable main sequence stars, which have a
multiplicity rate of 37$\pm$5\%.  We find that only 3\% of the cool
subdwarfs surveyed have companions within 10 AU, 3\% have companions
between 10 and 100 AU, and 14\% have companions beyond 100 AU. The
other 6\% of cool subdwarfs are spectroscopic binaries. This is very
different from K/M dwarfs that have most companions (13\%) at
separations closer than 10 AU.  However, because a search for close
binaries among a large sample of nearby cool subdwarfs remains
elusive, it is not yet settled whether or not the multiplicity rates
are significantly different.  Nonetheless, several different
observational results and theories pointing to a possible dearth of
subdwarf multiples are discussed.

\end{abstract}


\section{Introduction}

Studying stellar multiplicity is key to understanding stellar
formation, evolution, the true luminosity function of stellar
objects, and is relevant to our understanding of stellar and planetary
system stability.  Many binary surveys have been carried out on varied
populations, including the nearby Taurus star forming region
\citep{Leinert1993}, the Orion Nebula Cluster \citep{Kohler2006}, and
Herbig Ae/Be Stars \citep{Cordero2006}, to name a few.

The multiplicity of various samples of main sequence dwarfs has also
been discussed in the last decade.  \citet{Mason1998} found that more
than 59\% of O-type stars in clusters and associations have a visual,
speckle, or spectroscopic companion. \citet[~hereafter,
DM91]{Duquennoy1991} found that 57\% of solar-type binaries have
mass ratios greater than 0.1 after considering their survey
incompleteness. \cite{Henry1990} and \cite{Fischer1992} found that the
multiplicity fraction of M dwarfs drops to 34--42\%.  Thus, the
overall trend is that the multiplicity rate of main sequence stars
decreases with mass.

Subdwarfs are lower luminosity cousins of main sequence stars on the
HR diagram.  They are usually low metallicity stars and are often
referred to as either Population II or thick disk stars, making them
astrophysically distinct from dwarfs.  Given their formation in lower
metallicity environments, they might provide clues to the star
formation process if we could know (a) their multiplicity fraction and
(b) whether or not their multiplicity decreases with temperature, as
it does for their dwarf counterparts.  In this paper, we present
results from our optical speckle survey for cool subdwarf companions.
We then combine our results with other published subdwarf companion
surveys spanning G to M types, and we discuss our current
understanding of the multiplicity discrepancy between dwarfs and
subdwarfs.

\section{Sample Selection}

Our targets were selected from several different efforts, including
lists of spectroscopically identified subdwarfs \citep{Ryan1991,
Gizis1997, Jao2008} and stars with metallicity measurements
\citep{Carney1994, Cayrel2001, Nordstrom2004}.  In total, we selected
124 potential systems of interest, including 62 confirmed K and M type
subdwarf systems and 62 others of earlier type. Of the 124 subdwarf
systems, 118 have trigonometric distances within 60 pc \citep{YPC,
Hipparcos, Jao2005, Costa2005, Costa2006} while the remaining six
systems are beyond 60 pc (LHS 360, LHS 398, LHS 1481, LP907-080, HIP
101814 and HIP 101989). $V-K_{s}$ values greater than 2.0 are found
for 188 systems, while 6 other systems have $V-K_{s}$ less than 2.0 (G
048-039, G 062-044, G 083-034, G 140-046, G 141-008 and LHS 322).  The
62 subdwarfs within 60 pc have all been confirmed to have (a) K or M
types, (b) subdwarf spectroscopic features (strong CaH and TiO band
strength), and (c) [m/H] $\leqslant$$-$0.5 (if independent metallicity
measurements are available) or at least one magnitude below the fitted
main sequence line \citep{Jao2008}.  The 62 subdwarf systems are
listed in Table~\ref{tbl.speckle.observation}.

\section{Observations and Results}
\subsection{Observations and Calibrations of Speckle Interferometry}

Three observing runs were carried out at Kitt Peak National
Observatory (KPNO) and Cerro Tololo Inter-American Observatory (CTIO).
From 8--13 November 2005 and 1--9 August 2007, the United States Naval
Observatory (USNO) speckle camera \citep{Mason2006} was used on the
KPNO 4-m Mayall reflector.  The same USNO speckle camera was also used
on the CTIO 4-m Blanco telescope from 9--13 March 2006.  Observations
with this camera are made in or near $V$ band, usually with a
Str\"{o}mgren $y$ filter (5500$\pm$240\AA).

Speckle interferometry is a technique that is very sensitive to
changes in observing conditions, particularly when coherence length
($\rho_0$) and time ($\tau_0$) are degraded from nominal conditions.
Under nominal conditions the camera is capable of resolving companions
as close as 30 milliseconds of arc (mas) on these telescopes, provided
the companion has a small to moderate magnitude difference relative to
the primary star ($\Delta$m$\le$3). The $\Delta$m limit of the USNO
speckle camera during the three observing runs for this survey was
verified with nightly observations of pairs with known magnitude
differences \citep{Mason1996} with a wide range of separations. This
detection capability, both for large magnitude difference and
resolution close to the Rayleigh limit of the telescope, is
qualitatively established to be highly dependent on ambient seeing.
The faint limit of the USNO camera (V $\sim$ 16) was not approached
for the targets observed in this survey.

Calibration of the KPNO data was determined through the use of a
double-slit mask placed over the ``stove pipe'' of the 4-m telescope
during observations of a bright single star (as described in
\citealt{Hartkopf2000}).  This application of the well-known Young's
experiment allowed for the determination of scale and position angle
zero point without relying on binaries themselves to determine
calibration parameters.  Multiple observations through the slit mask
yield an error in the position angle zero point of 0\fdg11 and a
separation error of 0.17\%.  These ``internal errors" are undoubtedly
underestimates of the true errors of these observations, which we
anticipate are no larger than 0.5$\degr$ and 0.5\% in separation.

Because the slit-mask option is not available on the CTIO 4-m
telescope, we calibrated the southern hemisphere data using
observations of numerous wide equatorial binaries obtained at both the
KPNO and CTIO telescopes.  The calibration errors for these southern
observations were somewhat higher in position angle than those
achieved using the slit mask.  After removal of outliers, observations
of 47 pairs, including subdwarfs, O, B, G and M type stars, in common
to both telescopes yielded a zero-point error in $\theta$ of 0\fdg67
and a separation error of 1.44\%.  A small part of this error may
be attributed to such effects as orbital motion of pairs between
epochs of observation.  Regardless, the errors are rather larger than
desired, so more calibrations and quality control systems are in
preparation.

To verify companion detection limits a variety of known pairs with
reliable measures of differential magnitude and well-determined
ephemeris were observed.  This allows us to map out detection space
for both magnitude difference and resolution limit.  For stars that
were observed as single we can state with high confidence that in the
regime 30 mas $<$ $\rho$ $<$ 1\arcsec~these stars have no companions
with a magnitude difference of 3.0 or less.  This regime is relaxed to
50 mas $<$ $\rho$ $<$ 1\arcsec~for systems observed with a wider
filter bandpass (see below).  In total, 124 systems have been observed
using 130 total pointings, as listed in
Table~\ref{tbl.speckle.observation} (in six cases, wide companions
were observed with separate pointings).  In Table 1, columns (1--3)
are coordinates and star identifiers, column (4) indicates the
observing run (run epochs are given in the table notes), and columns
(5--9) give details of multiplicity.  Seven targets have also been
observed by \cite{Gizis2000} using Hubble Space Telescope (HST)
Wide-Field Planetary Camera 2 (WFPC2) and by \cite{Riaz2008} using HST
Advanced Camera for Survey (ACS); these are noted as ``W'' = WFPC2 or
``A'' = ACS in column (9) for the HST instruments used to make the
observations.

\subsection{Wide Companion Search}

In addition to our optical speckle efforts, we have also blinked
digital scans of photographic plates to reveal any wide common proper
motion pairs.  Images 10$\arcmin$$\times$10$\arcmin$ on a side were
extracted for each field from the Digitalized Sky Survey
(http://archive.stsci.edu/cgi-bin/dss\_form) for two epochs
corresponding to POSS-I red and POSS-II UKSTU red plates.  No new
companions were found through this effort.  We have also crosschecked
our target list on the L{\'e}pine Shara Proper Motion-North
(LSPM-North) catalog to search for any recently discovered companions
detected in that work.  The search limits of LSPM-North are fully
discussed in \cite{Lepine2005}, and all the wide companions are listed
on LSPM-North catalog.

\subsection{Results}

Table~\ref{tbl.speckle.observation} presents the complete list of
targets observed during the three observing runs.  The targets are
sorted into four different categories: (1) confirmed K and M subdwarfs
within 60 pc, (2) confirmed K and M subdwarfs beyond 60 pc, (3) G or
earlier than G type subdwarfs, and (4) probable, but unconfirmed by
spectroscopy, subdwarfs.  Most are null detections.  In some cases,
known companions were not detected because the binary separation is
larger than the speckle field of view (3\arcsec$\times$3\arcsec) or
the magnitude difference is larger than detectable with the optical
speckle camera.  For $\sim$40\% of the targets, an interference filter
with a significantly larger FWHM was used to allow enough signal
through the system for a companion search, at some cost in resolving
the closest pairs.  Specific stars for which the wider-band filter was
used are noted.  All of the observations are also available in the
Fourth Interferometic Catalog \citep{Hartkopf2001}\footnote{\tt
http://ad.usno.navy.mil/wds/int4.html}, which is updated frequently
on-line.

We found two new companions (LHS0182B and G125-026B) and confirm two
known companions (LHS0189B and G062-044B) and they are also listed in
Table~\ref{tbl.new}. The remaining 120 systems, including previously
known double-line and single-line spectroscopic binaries
\citep{Latham2002, Goldberg2002} and HST (ACS/WFPC2) targets, have no
companions detected by optical speckle interferometry.
Table~\ref{tbl.new} lists the astrometric measures of the four
detected companions, where columns (1--3) identify the systems by
providing the epoch-2000 coordinates, discovery designations (WSI =
Washington Speckle Interferometry) and alternate designations, and
columns (4--8) give the epoch of observation (expressed as a
fractional Besselian year), the position angle (in degrees), the
separation (in arcsec), and magnitude difference.  Note that the
position angles have not been corrected for precession, and are thus
based on the equinox for the epoch of observation.  The differential
magnitudes were determined by direct comparison of other pairs with
known magnitude differences and are probably accurate to
$\pm$0.5$_{\rm mag}$.

{\bf LHS 182} is type M0.0VI \citep{Gizis1997} at 43.3 pc \citep{YPC}.
The newly resolved companion with $\rho=$0\farcs62 lies at a projected
separation of 27 AU.

{\bf LHS0189/LHS0190} is a common proper motion binary and their
combined spectral type is a M3.0VI \citep{Jao2008}.  The 2\farcs81 projected
separation is equivalent to 62 AU at a distance of 22.1 pc
\citep{Costa2006}.  Because this system is fainter than G 125-026
(discussed below), no speckle work was done on individual components.

{\bf G062-044} is a G type subdwarf with [m/H]$=$$-$0.69
\citep{Carney1994} and is a single-line spectroscopic binary
\citep{Latham2002} with a period of 3.3 years. \cite{Balega2006} used
their speckle camera on the 6-m telescope of the Special Astrophysical
Observatory to resolve this system with $\rho=$0\farcs082 at
62.5$^{\circ}$ in 2001.271. Our data show the companion at
$\rho=$0\farcs13 and 147.6$^{\circ}$ in 2006.192, implying significant
orbital motion.

\cite{Carney1994} reported {\bf G125-026} to have [m/H]$=$$-$1.5,
while its $V-K=$2.5 indicates it is a mid-K type subdwarf.  Because
this system has a wide separation (3\farcs29 along the diagonal of our
field of view, corresponding to a projected separation of 181 AU at
55.2 pc), both components were observed.  No companion was detected
for either component.  However, G125-026 is in a crowded field, so the
companion is possibly optical.  A follow-up observation is necessary
to confirm common proper motions.

\section{Multiplicity Comparison with Similar Surveys}

Our sample is a mixture of confirmed K/M subdwarfs, G subdwarfs, and
subdwarf candidates.  Here we use the 62 systems in the confirmed K/M
type subdwarf category as the benchmark sample for our multiplicity
discussion.  For these 62 systems the (single:double:triple:quadruple)
ratios are 46:12:2:2.  Hence, the fraction of K/M type subdwarf
multiple systems for this sample is 26$\pm$6\%.

A few recent subdwarf companion surveys have also yielded relatively
low multiplicity fractions.  \cite{Gizis2000} used HST/WFPC2 to
observe eleven cool subdwarfs and failed to detect any companions.  We
observed five of those targets, and also detected no companions.
\cite{Lepine2007} observed 18 subdwarfs from Lick Observatory using
the AO-Laser Guide Star system and resolved only one system, LSR
1530$+$5608.  Recently, \cite{Riaz2008} reported no companions
detected by HST/ACS for 19 M type subdwarfs, other than the known wide
common proper motion system LHS2139/2140 ($\rho=$6\farcs2).  We
observed two of those targets, LHS 161 and LHS 482, and also detected
no companions.  However, Gelino (2007, private communication) used the
Keck AO-Natural Guide Star system to observe 54 low metallicity stars
of spectral type from G to M, and detected seven possible companions,
five of which are new. Recently, a late-type subdwarf, LSR 1610$-$00
(\citealt{Dahn2008}, $V-I=$4.05, and d=32.5 pc), is discovered to be a
binay through parallax observations, not through a high resolution
companion survey.

The binary rate from the combined samples of \cite{Gizis2000},
\cite{Lepine2007}, \cite{Riaz2008} and \cite{Dahn2008}, which had no
stars in common, is only 6\% (3/49).  Gelino's work yields a higher
binary rate of 13\%, but the sample is a mixture of G, K and M type
subdwarfs, so the results are not immediately comparable to the other
surveys.  Our survey is the largest available to date, and so far is
the only survey for which all targets have trigonometric parallaxes,
in a volume-limited sample reaching to 60 pc (details in the next
paper of this series).

\section{Multiplicity Comparison with K and M type Dwarfs}

Binary surveys have been carried out for M dwarfs, the main sequence
counterparts of the stars surveyed here, and comparisons between dwarf
and subdwarf multiplicities are now possible.  \cite{Henry1990} used
an infrared speckle camera to survey for stellar and brown dwarf
companions to 27 known M dwarfs within 5 pc north of DEC = $-$30$\degr$, and
found a multiplicity fraction of 34$\pm$9\%.  \cite{Henry1991}
expanded the sample to 99 M dwarfs within 8 pc north of DEC = $-$25$\degr$,
and found a consistent multiplicity fraction of 31$\pm$6\%.
\cite{Fischer1992} did a complete analysis of companion searches
around 100 M dwarfs within 20 pc, including radial velocity, visual
(astrometric), infrared imaging, and the infrared speckle efforts.
They found a multiplicity fraction of 42$\pm$9\% after considering the
incompleteness of the surveys.  \cite{Reid1997} (hereafter, RG97)
compiled a sample of 106 low mass stars (80\% of which were M dwarfs)
north of DEC = $-$30$\degr$ and found a multiplicity fraction of
35$\pm$5\%.  More recently, \cite{Delfosse2004} observed 100 M dwarfs
within 9 pc using radial velocity and AO observations, and found a
multiplicity fraction of 26$\pm$3\%.  Many of the studies include the
same stars, and all reach the same conclusion --- the multiplicity of
M dwarfs is roughly 30--40\%.

Here we compare our results to the work of RG97.  In order to match
our sample of K/M subdwarfs, non-K/M type dwarfs are excluded from
RG97, leaving 92 stars.  Of these, 58 are single and 34 are multiples
(counting only companions with stellar masses, i.e., masses greater
than 0.08M$_{\odot}$), yielding a multiplicity fraction of 37$\pm$5\%.
Hence, the multiplicity rate difference between cool dwarfs and
subdwarfs is 11\%.

Table~\ref{tbl.our.RG97} compares the techniques used for our sample
and RG97's sample.  All of our subdwarf targets have been searched
using the optical speckle camera, which will detect companions in the separation 
regime 50 mas $<$ $\rho$ $<$ 1\arcsec~with a magnitude
differences of 3.0 or less, and by blinking photographic plates for
wide common proper motion companions.  Although 43 targets in our
sample have radial velocity measurements, only six systems have been
observed for spectroscopic binaries using radial velocity surveys, so
spectroscopic companions are almost certainly underrepresented.  Only
two other stars (LHS 64 and LHS 482) in
Table~\ref{tbl.speckle.observation} are reported to be spectroscopic
binaries \citep{Dawson2005}, but no orbital elements are
available.\footnote{\cite{Dawson2005} reported LHS 169 has a large
radial velocity variation, so it is flagged as a ``possible''
spectroscopic binary by them; we count it as a binary in this
analysis.}

Among the 92 systems selected from the RG97 work, 68 were observed by
\cite{Henry1991} using an infrared speckle camera, which detects
companions in the regime 0.2\arcsec~$<$ $\rho$ $<$ 2.0\arcsec~on the
Steward Observatory 90in telescope used for the observations at 2.2
$\mu$m~.  In addition, 36 systems were observed by \cite{Marcy1989}
and \cite{Delfosse1999} during long term radial velocity surveys
covering orbital periods of a few days to a few years.  More details
about the companion search for this sample are discussed in
RG97. Consequently, many of the 92 stars in our comparison sample have
been searched for companions with separations of a few AU to thousands
of AU.

The top two plots of Figure~\ref{fig.subdwarf.dwarf} illustrate the
distribution of companion separations from the dwarf and subdwarfs
samples.  It appears that cool subdwarf binaries tend to have larger
separations.  However, as discussed above, our subdwarf sample has not
yet been systematically searched for close companions via long-term
radial velocity surveys.  Note that the optical and infrared speckle
efforts search similar spatial regimes --- the factor of roughly five
difference in resolution limit is compensated for because the dwarfs
are closer than the subdwarfs by about a factor of five.  Although the
dearth of close binaries in our sample may be due to lack of
observational coverage, \cite{Abt2008} offered a possible scenario to
explain why metal-poor stars might have lower multiplicity fractions
than more metal-rich stars.  According to $n$-body simulations,
binaries become tighter if they survive interactions in dense
clusters.  This implies that the metal-poor field subdwarfs we see
today may have had shorter lifetimes in clusters than generally
younger, more metal-rich stars.  Clearly, a subdwarf radial velocity
survey of K/M subdwarfs needs to be done to confirm or refute our
tentative conclusion that subdwarfs have fewer binaries.

The bottom of Figure~\ref{fig.subdwarf.dwarf} plots the luminosity
functions (LFs) from both surveys to see if there are any significant
differences.  The LF of K/M subdwarfs has two prominent peaks, in the
$M_{Ks}=$4--5 bin and in the $M_{Ks}=$7--8 bin.  In contrast, the LF
of dwarfs looks more like a normal distribution.  There are several
reasons for these differences.  First, our sample lacks K type
subdwarfs.  As discussed in \cite{Jao2008}, K type subdwarfs are
difficult to separate spectroscopically from dwarfs using low
resolution spectra covering from 6000\AA~to 9000\AA~(the region widely
used to identify cool subdwarfs).  Consequently, our current K type
subdwarfs within 60 pc are undersampled.  Second, most of the
subdwarfs with $M_{K_{s}}<$6.0 in our sample were selected
spectroscopically from \cite{Carney1994}, so few subdwarfs with
$M_{K_{s}}$ = 5--6 were included in our survey.  Third, very few
subdwarfs fainter than $M_{K_{s}}=$9 within 60 pc are yet known.
Thus, the LFs of the two samples appears different.  Will subdwarf
multiplicity be the same as that of dwarfs if we increase the 60 pc
sample to smooth out the LF, and we expand our companion searches?
While only future efforts can answer this question for K/M stars, we
can make some comparisons now by discussing the multiplicity fractions
of A to G type stars for both dwarf and subdwarf samples.

\section{Discussion}


The multiplicity of G type subdwarfs seems to be different from that
of their main sequence counterparts.  Four different surveys are
discussed below: 1) \cite{Stryker1985} found a lower limit of 20\% to
30\% for the binary frequency of their subdwarfs from radial velocity
work, which are primarily spectral types A, F and G type subdwarfs. 2)
\cite{Zapatero2004} argued that only 15\% of the metal-poor (G to
early M) stars have stellar companions as close as 1\farcs67 using CCD
images. 3) \cite{Zinnecker2004} used optical speckle interferometry,
AO and direct imaging methods and found the binary fraction of the
combined Carney-Latham \citep{Carney1994} and \cite{Norris1986}
samples to be up to 15\%, depending on the flux ratio
($F_{primary}$/$F_{secondary}$) cutoff assigned.  Again, most of their
targets were AFG subdwarfs as well. 4) \cite{Rastegaev2007} observed
106 systems from the Carney-Latham sample also using optical speckle,
and they found their multiple star fraction to be 33\% (75 singles and
35 multiples).  Based on these four surveys, the multiplicity rate is
between 15\% and 33\%.

Table~\ref{tbl:fraction} outlines the differences noted to date
between main sequence and subdwarf multiplicities. From this table, 
there are two ways to interpret the results. We
can either compare the multiplicity rate (horizontally) between A--G
and K \& M type stars, or we can compare the multiplicity difference
(vertically) between dwarfs and subdwarfs in each type. Before we
compare horizontally or vertically, we choose the multiplicity
fractions of dwarfs as reference frames. In other words, we assume
the multiplicity studies for dwarfs are comprehensive and
complete. Note that the multiplicity rate for later type stars {\it
should} be less for the early type stars, because of the decreased
companion mass range. The results of companion surveys of subdwarfs
should also follow this trend.

\begin{description}
\item[Horizontal Comparison:] Given that the quantity of M dwarf
multiples is roughly 65\% that of G dwarfs for main sequence stars, we
might expect either 1) A--G type subdwarfs should be as high as 40\%,
or 2) the K and M subdwarfs multiplicity fraction should be no larger
than 20\%, if we {\it assume} the binary formation theory is
independent of the metallicity. The first expectation implies that the
multiplicity rate of subdwarfs shown in the Table~\ref{tbl:fraction}
could be underestimated. The second expectation implies that our
volume-limited sample contains more binaries than it should, so more
single stars within 60 pc are missing from this sample. Because most
of these early type subdwarfs have been searched for companions from
close to wide separations, we think the multiplicity rate for them
will not be much different. However, we are sure that the sample
of K and M subdwarfs within 60 pc is incomplete, because a)
historically these stars have been neglected from trigonometric
parallax observations and b) we have not completed the optical speckle
survey of the entire cool subdwarfs within 60 pc\footnote{We continue
this optical speckle survey in 2008 at KPNO and CTIO and data are
still under analysis.}. Consequently, a census of nearby K and M
subdwarf sample is necessary to understand this horizontal
comparison, and we are continuing to make subdwarf parallax
observations through the Cerro Tololo Inter-american Observatory
Parallax Investigation (CTIOPI) project.

\item[Vertical Comparison:] It appears that the multiplicity rate for
subdwarfs is less than for dwarfs, but the differences are not equal
between early and late type stars. For A--G type stars, the subdwarf 
multiplicity rate is only about 50\% of dwarfs. For K and M type
stars, the subdwarf multiplicity rate is about 34\% that of dwarf. This
shows they are different, but recent studies do not all support this
trend.

\cite{Latham2002} found in their high proper motion sample, which
contains both old and young populations, that there is no significant
difference in the binary fraction of the two populations.
\cite{Grether2007} argued stellar companions tend to be more abundant
around low metallicity hosts (see their Figure 15) when they examined
the relation between frequency close companions and the metallicity of
FGK stars.

Theoretical work is also in conflict with this
trend. \cite{Bate2005} has investigated how the metallicity of a
molecular cloud affects fragmentation.  By setting the critical
density of the equation of state a factor of 9 lower than normal
calculation (\citealt{Bate2003}), he tried to mimic the thermal
dynamics for molecular gas that has a lower metallicity.  He found
this does not affect the ability to form close binaries. However, this
low density (metallicity) calculation generates slightly more binaries
(17\%, 19 singles and 4 multiples) than the other calculation (13\%, 26
singles and 4 multiples).

\cite{Fischer2005} also found the number of stars with planets
decreases as a power law when primary stars decrease their metallicity
(see their Figure 5). If we assume there is some degree of similarity
between planet formation and binary formation as a function of
metallicity, fewer subdwarf companions are expected.

\end{description}

\begin{figure}
\centering
\includegraphics[scale=0.7, angle=90]{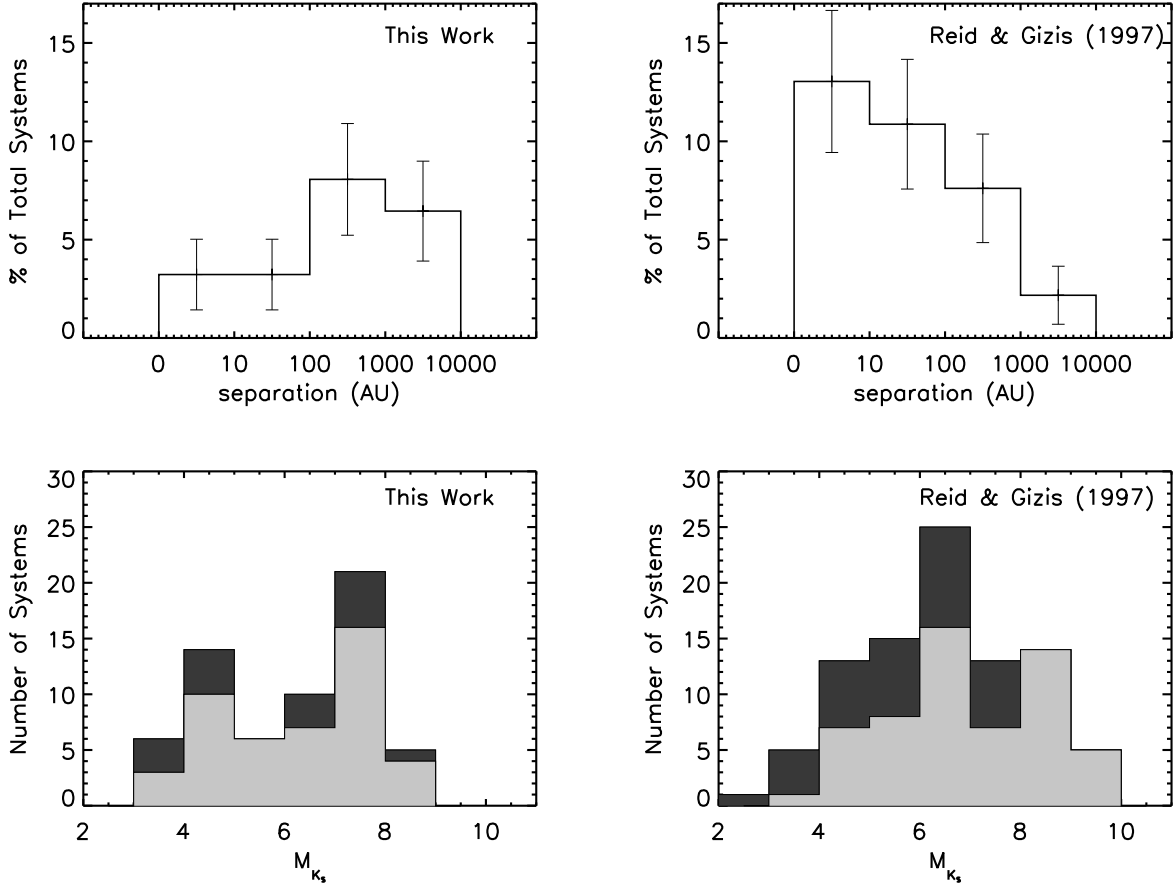}

\caption{Top: Comparison of the secondary projected separations of our
sample and RG97. Because two of the objects in our sample, G016$-$009
and G129$-$042, are SB2, they are plotted at their mininum separations
($a\sin i$, where $i=$90$^{\circ}$), as is also done for two SB2 (GJ
268 and GJ 829) in RG97. Three spectroscopic binaries in our sample
(LHS 64, LHS 482 and 169) and two in RG97 (G 041-014 and G 203-047) do
not have $a\sin i$ available in publications, so they are not shown in
these top two plots. Bottom: Comparison of the LFs of our sample and
RG97.  In this figure we plot the primary stars only. The black bars
indicate primary stars in binary systems and gray bars indicate single
star systems. Because SB2 systems do not have magnitude differences
available at K band, SB2 systems are plotted using combined
photometry, instead of using de-convolved single star photometry. For
systems with only $\Delta$$V$ available, we use the approximate
relation ($\Delta m_{K}$/$\Delta m_{V}$=0.53) discussed in
\cite{Probst1981} to convert it to $\Delta K$.}
\label{fig.subdwarf.dwarf}
\end{figure}

Apparently, these multiplicity comparisons are still unsettled, no
matter if they are compared horizontally or vertically in
Table~\ref{tbl:fraction}. Consequently, the following investigations
are necessary to better understand the multiplicity differences
between cool dwarfs and subdwarfs:

\begin{itemize}

\item As we discussed above, increasing the total number of nearby
cool subdwarfs is one of the keys to understanding their multiplicity. We
can select these subdwarfs through reduced-proper motion diagram, then
observe spectroscopically to confirm their luminosity classes. 
Parallax observations can then be carried out to secure their distances.

\item \cite{Abt2008} concluded that the multiplicity discrepancy
between metal-poor and metal-rich stars depends mainly on the
equipment/techniques used. He said using high spectral resolution
spectra for finding spectroscopic binaries tends to conclude there is
no difference in the binary frequency, but while using low spectral
resolution spectra in radial velocity survey concludes that the
metal-pool stars have fewer binary than metal-rich stars.  As
Figure~\ref{fig.subdwarf.dwarf} shows, currently there are only two
cool subdwarf companions (2/16$\approx$13\%) within 10 AU and both of
them are plotted in projected minimum separations ($i$=90$^{\circ}$).
Hence, {\it these K and M subdwarfs need to have a completed radial
velocity survey to detect any possible close binaries}.

On the other hand, \cite{Chaname2004} and \cite{Lepine2005} have
reported many wide common proper motion (CPM) subdwarf binaries on the
reduced-proper motion diagram. Unfortunately, all of these CPM
binaries need to have spectroscopic follow-up to confirm their
luminosity classes, and need parallax observations. If the radial
velocity surveys for cool subdwarfs fail to detect companions closer
than 10 AU, it will give an important observational support to the
theory that metal-poor stars lack short-period binaries.

\item \cite{Tokovinin2004} concluded that systems of multiplicity three and
higher are frequent, accounting for about 20\% of the total population
of stellar systems, so forming hierarchical multiples is not a rare
phenomenon.  Currently, our sample contains four systems (6\%) with
three or more components. Among these four systems, there is only one
system with $V-K_{s}>$2.6, and the rest of these systems are K type
subdwarfs.  Supposedly, those hierarchical multiples of M type
subdwarfs should exist, but have not been identified.

\end{itemize}

\begin{acknowledgements}

This work has been supported at GSU by NASA's Space Interferometry
Mission, the National Science Foundation (NSF, grant AST-0507711), and
GSU. This research has made use of the SIMBAD database, operated at
CDS, Strasbourg, France. This work also has used data products from
the Two Micron All Sky Survey, which is a joint project of the
University of Massachusetts and the Infrared Processing and Analysis
Center at California Institute of Technology funded by NASA and NSF.

\end{acknowledgements}



\begin{deluxetable}{ccclclccl}
\tabletypesize{\small}
\tablewidth{18cm}
\tablecaption{Speckle Observations}
\tablehead{
\colhead{RA}        &
\colhead{DEC}       &
\colhead{Name}      &
\colhead{Site}      &
\multicolumn{5}{c}{Binary Note} \\
                     &
                     &
                     &
                     &
CPM Comp.            &
$\rho$(\arcsec)      &
$\theta$($^{\circ}$)    &
epoch (yrs)          &
Others               \\
(1)                  &
(2)		     &
(3)		     &
(4)		     &
(5)		     &
(6)		     &
(7)                  &
(8)                  &
(9)}
\startdata
\multicolumn{8}{c}{Confirmed K and M subdwarfs within 60 pc} \\
\tableline
00 09 16.46 &  $+$09 00 41.9 & LHS0104            &      K1\tablenotemark{a}  &                   &             &         &          &                                            \\
00 12 30.31 &  $+$14 33 48.8 & G030-052           &      K2                   &                   &             &         &          &                                            \\
00 17 40.01 &  $-$10 46 16.9 & LHS0109            &      K2\tablenotemark{a}  &                   &             &         &          &                                            \\
00 45 16.67 &  $+$01 40 34.4 & G001-021B          &      K1\tablenotemark{a}  &    G001-021A      &  19.1       &  242.6  & 2000.60  &                                            \\
00 45 17.80 &  $+$01 40 43.3 & G001-021A          &      K1\tablenotemark{a}  &                   &             &         &          &  SB1                                       \\
02 02 52.16 &  $+$05 42 21.0 & LHS0012            &      K1\tablenotemark{a}  &                   &             &         &          &                                            \\
02 08 23.89 &  $+$28 18 18.3 & G072-059           &      K1                   &    G072-058       &  210.7      &  181.1  & 2006.89  &  SB1                                       \\
02 08 23.91 &  $+$28 18 38.9 & G072-058           &      K1                   &                   &             &         &          &  SB1                                       \\
02 34 12.46 &  $+$17 45 50.5 & LHS0156            &      K2\tablenotemark{a}  &                   &             &         &          &                                            \\
02 52 45.51 &  $+$01 55 50.5 & LHS0161            &      K1\tablenotemark{a}  &                   &             &         &          &  A                                         \\
03 06 28.67 &  $-$07 40 41.5 & LHS0165            &      K1\tablenotemark{a}  &                   &             &         &          &                                            \\
03 13 24.24 &  $+$18 49 37.7 & LHS0169            &      K1\tablenotemark{a}  &                   &             &         &          &  W, SB?                                    \\
03 16 26.81 &  $+$38 05 55.8 & LHS0170            &      K1                   &                   &             &         &          &                                            \\
03 19 39.85 &  $+$33 35 55.0 & G037-034           &      K1                   &                   &             &         &          &                                            \\
03 28 53.13 &  $+$37 22 56.7 & LHS0173            &      K1                   &                   &             &         &          &                                            \\
03 30 44.82 &  $+$34 01 07.2 & LHS0174            &      K1                   &                   &             &         &          &  W                                         \\
03 38 15.70 &  $-$11 29 13.5 & LHS0020            &      K1\tablenotemark{a}  &                   &             &         &          &                                            \\
03 42 29.45 &  $+$12 31 33.8 & LHS0178            &      K1\tablenotemark{a}  &                   &             &         &          &                                            \\
03 47 02.11 &  $+$41 25 38.2 & LHS0180            &      K2                   &                   &             &         &          &                                            \\
03 47 02.63 &  $+$41 25 42.4 & LHS0181            &      K1, K2               &     LHS0180       &  7.4        &  55.3   & 2003.91  &  SB1                                       \\
03 50 13.89 &  $+$43 25 40.5 & LHS0182            &      K1                   &    \multicolumn{5}{c}{Resolved by speckle and see Table~\ref{tbl.new}}                            \\
04 03 15.00 &  $+$35 16 23.8 & LHS0021            &      K1                   &                   &             &         &          &                                            \\
04 03 38.44 &  $-$05 08 05.4 & LHS0186            &      K1\tablenotemark{a}  &                   &             &         &          &                                            \\
04 04 20.30 &  $-$04 39 18.3 & HD025673           &      C1                   &                   &             &         &          &                                            \\
04 25 38.35 &  $-$06 52 37.0 & LHS0189/0190       &      K1                   &    \multicolumn{5}{c}{Resolved by speckle and see Table~\ref{tbl.new}}                 \\
04 25 46.76 &  $+$05 16 03.0 & G082-018           &      K1\tablenotemark{a}  &                   &             &         &          &                                            \\
05 11 40.60 &  $-$45 01 06.4 & Kapteyn            &      C1                   &                   &             &         &          &                                            \\
06 22 38.57 &  $-$12 53 05.1 & LHS1841            &      C1\tablenotemark{a}  &                   &             &         &          &                                            \\
06 44 42.97 &  $+$14 54 36.0 & HIP32308           &      K1                   &                   &             &         &          &                                            \\
07 02 36.44 &  $+$31 33 54.7 & G087-019           &      K1                   &                   &             &         &          &                                            \\
09 43 46.16 &  $-$17 47 06.2 & LHS0272            &      K1\tablenotemark{a}  &                   &             &         &          &                                            \\
10 13 01.62 &  $-$39 06 07.9 & LTT3743            &      C1\tablenotemark{a}  &                   &             &         &          &                                            \\
11 10 02.64 &  $-$02 47 26.4 & G010-003           &      C1\tablenotemark{a}  &                   &             &         &          &                                            \\
11 11 13.68 &  $-$41 05 32.7 & LHS0300A           &      C1\tablenotemark{a,b}&     LHS0300B      &   4.3       &  62.0   &  2001.   &                                            \\
11 52 58.73 &  $+$37 43 07.3 & LHS0044            &      K1                   &                   &             &         &          &                                            \\
11 58 28.02 &  $-$41 55 19.3 & LHS2485            &      C1\tablenotemark{b}  &     LHS2484       &   23.6      &  312.4  &  2000.20 &                                            \\
12 02 33.66 &  $+$08 25 50.7 & LHS0320            &      C1\tablenotemark{a}  &                   &             &         &          & W                                          \\
12 24 26.81 &  $-$04 43 36.7 & LHS0326            &      C1\tablenotemark{a}  &                   &             &         &          &                                            \\
12 56 23.74 &  $+$15 41 44.5 & LHS0343            &      C1\tablenotemark{a}  &                   &             &         &          &                                            \\
13 18 56.71 &  $-$03 04 17.9 & LHS2715            &      C1\tablenotemark{a}  &                   &             &         &          &                                            \\
14 02 46.66 &  $-$24 31 49.6 & LHS2852            &      C1\tablenotemark{a}  &                   &             &         &          &                                            \\
15 10 12.96 &  $-$16 27 46.6 & LHS0052            &      C1                   &      LHS0053      &  300.7      &  180.3  &  2000.   &                                            \\
15 10 13.08 &  $-$16 22 46.0 & LHS0053            &      C1                   &                   &             &         &          &                                            \\
15 28 13.99 &  $+$16 43 10.8 & LHS3073            &      K2\tablenotemark{a}  &                   &             &         &          &                                            \\
15 34 40.11 &  $+$02 12 15.1 & LHS3084            &      C1\tablenotemark{a}  &                   &             &         &          &                                            \\
15 43 18.33 &  $-$20 15 32.9 & LHS0406            &      C1\tablenotemark{a}  &                   &             &         &          &                                            \\
15 45 52.41 &  $+$05 02 26.6 & G016-009           &      K2                   &                   &             &         &          &  SB2                                       \\
16 08 55.40 &  $+$01 51 07.5 & G016-031           &      C1\tablenotemark{a}  &                   &             &         &          &                                            \\
16 20 17.97 &  $-$48 13 32.8 & LHS3182            &      C1                   &                   &             &         &          &                                            \\
16 37 05.42 &  $-$01 32 00.5 & LHS0424            &      C1\tablenotemark{a}  &                   &             &         &          &                                            \\
16 42 04.33 &  $+$10 25 58.7 & LHS0425            &      C1\tablenotemark{a}  &                   &             &         &          &                                            \\
18 41 36.37 &  $+$00 55 13.8 & LHS0467            &      K2\tablenotemark{a}  &                   &             &         &          &                                            \\
18 45 52.24 &  $+$52 27 40.6 & LHS3409            &      K2\tablenotemark{a}  &                   &             &         &          & W                                          \\
19 07 02.04 &  $+$07 36 57.3 & G022-015           &      K2\tablenotemark{a}  &                   &             &         &          &                                            \\
19 19 00.52 &  $+$41 38 04.5 & G125-004           &      K1                   &                   &             &         &          &                                            \\
19 39 57.41 &  $+$42 55 57.0 & G125-026A          &      K2\tablenotemark{a}  &    \multicolumn{5}{c}{Resolved by speckle and see Table~\ref{tbl.new}}                 \\
19 39 57.41 &  $+$42 55 57.0 & G125-026B          &      K2\tablenotemark{a}  &                   &             &         &          &                                            \\
20 05 02.20 &  $+$54 26 03.2 & LHS0482            &      K2\tablenotemark{a}  &                   &             &         &          & SB                                         \\
20 27 29.09 &  $+$35 59 24.8 & LHS0491            &      K2\tablenotemark{a}  &                   &             &         &          & A                                          \\
21 07 55.39 &  $+$59 43 19.4 & LHS0064            &      K2\tablenotemark{a}  &                   &             &         &          & SB                                         \\
21 32 11.93 &  $+$00 13 18.0 & G026-009ACD        &      K2                   &                   &             &         &          & SB2                                        \\
21 32 16.22 &  $+$00 15 14.4 & G026-010           &      K2\tablenotemark{a}  &     G026-009ACD   &  133.1      &   29.3  &   2003.  &                                            \\
22 14 24.01 &  $-$08 44 42.0 & LHS3780            &      K2\tablenotemark{a}  &                   &             &         &          &                                            \\
22 31 47.83 &  $+$49 42 13.5 & G215-053           &      K2                   &                   &             &         &          &                                            \\
23 08 26.04 &  $+$31 40 23.9 & LHS0536            &      K1\tablenotemark{a}  &                   &             &         &          & W                                          \\
23 25 11.31 &  $+$34 17 14.0 & LHS3942            &      K2\tablenotemark{a}  &                   &             &         &          &                                            \\
23 43 16.74 &  $-$24 11 16.4 & LHS0073            &      K2\tablenotemark{a}  &      LHS0072      &  94.3       &   153.4 &   2000.  &                                            \\
23 55 04.17 &  $+$20 23 05.5 & G129-042           &      K1                   &                   &             &         &          & SB2                                        \\
\tableline
\multicolumn{8}{c}{Confirmed K and M subdwarfs beyond 60 pc} \\
\tableline						       
02 58 10.24 &  $-$12 53 05.9 & LHS1481            &      K1\tablenotemark{a}  &                   &             &         &          &                                            \\
13 46 55.52 &  $+$05 42 56.4 & LHS0360            &      C1\tablenotemark{a}  &                   &             &         &          &                                            \\
15 34 27.75 &  $+$02 16 47.5 & LHS0398            &      C1\tablenotemark{a}  &                   &             &         &          &                                            \\
\tableline						       
\multicolumn{8}{c}{G or earlier than G type subdwarfs} \\	       
\tableline						       
01 04 26.46 &  $-$02 21 59.8 & G070-035           &      K1                   &                   &             &         &          &                                            \\
02 25 49.75 &  $+$05 53 39.5 & G073-056           &      K1\tablenotemark{a}  &                   &             &         &          &                                            \\
07 54 34.19 &  $-$01 24 44.3 & G112-054           &      K1, C1               &                   &             &         &          & SB1                                        \\
12 06 00.94 &  $+$14 38 56.8 & G012-016           &      C1                   &                   &             &         &          &                                            \\
13 31 39.95 &  $-$02 19 02.5 & G062-044           &      C1                   &     \multicolumn{5}{c}{Resolved by speckle and see Table~\ref{tbl.new}}                \\
16 13 48.56 &  $-$57 34 13.8 & LHS0413            &      C1                   &                   &             &         &          &                                            \\
18 12 21.88 &  $+$05 24 04.5 & G140-046           &      K1                   &                   &             &         &          &                                            \\
20 32 51.67 &  $+$41 53 54.7 & G209-035           &      K1                   &                   &             &         &          & SB2                                        \\
\tableline						       
\multicolumn{8}{c}{Possible Subdwarfs} \\		       
\tableline						       
00 12 46.96 &  $+$54 39 45.4 & LHS1039            &      K1\tablenotemark{a}  &                   &             &         &          &                                            \\
00 38 29.17 &  $+$42 59 59.8 & HIP003022A         &      K1                   &     HIP003022B    &   53.1      &  124.0  &  2000.   &                                            \\
00 40 49.27 &  $+$40 11 13.8 & HD003765           &      K1                   &                   &             &         &          &                                            \\
00 49 34.47 &  $+$97 57 09.6 & G243-041           &      K1\tablenotemark{a}  &                   &             &         &          &                                            \\
01 18 41.07 &  $-$00 52 03.0 & G070-051           &      K1                   &     G070-050      &   27.9      &  208.5  &  1998.71 &                                            \\
01 38 14.19 &  $+$17 49 45.9 & G003-010           &      K1                   &                   &             &         &          &                                            \\
03 23 33.48 &  $+$43 57 26.2 & HIP015797          &      K1                   &                   &             &         &          &                                            \\
03 26 04.26 &  $+$45 27 28.4 & HIP015998          &      K1                   &                   &             &         &          &                                            \\
03 43 55.34 &  $-$19 06 39.2 & HD023356           &      K1                   &                   &             &         &          &                                            \\
03 55 03.80 &  $+$61 10 00.6 & G246-053           &      K1                   &                   &             &         &          &                                            \\
05 54 04.24 &  $-$60 01 24.5 & HD040307           &      C1                   &                   &             &         &          &                                            \\
06 06 03.51 &  $-$59 32 35.3 & LHS1818            &      C1                   &                   &             &         &          &                                            \\
06 06 24.66 &  $+$63 50 06.6 & G249-037           &      K1                   &                   &             &         &          &                                            \\
06 58 27.86 &  $+$18 59 49.7 & G088-001           &      K1                   &                   &             &         &          &                                            \\
07 08 04.23 &  $+$29 50 04.1 & HD053927           &      K1                   &                   &             &         &          &                                            \\
07 58 04.37 &  $-$25 37 35.8 & HD065486           &      C1                   &                   &             &         &          &                                            \\
08 04 34.65 &  $+$15 21 51.3 & G040-005           &      K1                   &                   &             &         &          &                                            \\
08 39 50.78 &  $+$11 31 21.4 & G009-013           &      K1, C1               &                   &             &         &          &                                            \\
08 40 33.55 &  $+$13 33 23.4 & G009-014           &      K1                   &                   &             &         &          &                                            \\
08 43 18.03 &  $-$38 52 56.5 & HD074576           &      C1                   &                   &             &         &          &                                            \\
09 00 47.41 &  $+$21 27 13.8 & G009-042           &      K1                   &                   &             &         &          &                                            \\
09 49 48.53 &  $+$11 06 22.9 & G048-039           &      K1                   &                   &             &         &          & SB1                                        \\
10 41 02.02 &  $+$03 35 46.6 & HIP052285A         &      K1                   &      HIP052285B   &   7.5       &  119.9  &  2004.37 &                                            \\
10 53 23.80 &  $+$09 44 21.8 & G044-045           &      K1                   &                   &             &         &          &                                            \\
11 58 27.19 &  $-$27 40 09.8 & LP907-080          &      C1\tablenotemark{a}  &                   &             &         &          &                                            \\
12 15 10.55 &  $-$10 18 44.9 & LHS0322            &      C1                   &                   &             &         &          &                                            \\
12 55 15.97 &  $+$07 49 57.6 & HIP63063           &      C1                   &                   &             &         &          &                                            \\
12 59 01.56 &  $-$09 50 02.7 & HIP063366          &      C1\tablenotemark{c}  &                   &             &         &          &                                            \\
13 16 51.05 &  $+$17 01 01.8 & GJ0505A            &      C1\tablenotemark{b}  &       GJ0505B     &   7.5       &  106.6  & 2004.28  &                                            \\
13 31 06.17 &  $-$04 06 20.0 & HIP065940          &      C1\tablenotemark{a}  &                   &             &         &          &                                            \\
13 52 35.87 &  $-$50 55 18.3 & HIP67742A          &      C1\tablenotemark{b}  &       HP67742B    &   5.8       &  82.6   & 1987.36  &                                            \\
13 57 18.14 &  $+$06 58 55.2 & HIP68165           &      C1\tablenotemark{a}  &                   &             &         &          &                                            \\
14 10 02.68 &  $-$61 31 18.5 & LHS2871            &      C1                   &                   &             &         &          &                                            \\
15 20 26.13 &  $+$00 14 40.7 & G015-017           &      C1                   &       G015-018    &   196.6     &  171.0  & 2000.31  &                                            \\
16 09 42.79 &  $-$56 26 42.5 & HD144628           &      C1                   &                   &             &         &          &                                            \\
18 09 37.41 &  $+$38 27 28.0 & HD166620           &      K1                   &                   &             &         &          &                                            \\
18 26 10.08 &  $+$08 46 39.3 & G141-008           &      K1                   &                   &             &         &          & SB1                                        \\
19 12 45.01 &  $+$18 48 45.6 & G142-015           &      K1                   &                   &             &         &          &                                            \\
19 37 14.10 &  $+$70 44 29.1 & G260-028           &      K1                   &                   &             &         &          &                                            \\
19 50 55.86 &  $+$03 56 48.3 & G023-012           &      K1                   &                   &             &         &          &                                            \\
19 58 35.40 &  $+$81 16 12.2 & HIP098322          &      K1                   &                   &             &         &          &                                            \\
20 03 52.12 &  $+$23 20 26.4 & HD190404           &      K1                   &                   &             &         &          &                                            \\
20 13 59.85 &  $-$00 52 00.7 & HD192263           &      K1                   &                   &             &         &          &                                            \\
20 37 58.49 &  $+$77 14 02.3 & HIP101814          &      K1                   &                   &             &         &          &                                            \\
20 40 07.90 &  $+$41 15 10.7 & HIP101989          &      K1                   &                   &             &         &          &                                            \\
22 35 13.31 &  $+$18 06 20.1 & G083-034           &      C1                   &                   &             &         &          & SB1                                        \\
23 13 16.98 &  $+$57 10 06.0 & HD219134           &      K1, K2               &                   &             &         &          &                                            \\
23 19 58.19 &  $+$28 52 03.9 & HD219953           &      K1                   &                   &             &         &          &                                            \\
23 35 49.27 &  $+$00 26 43.6 & G029-050           &      K1                   &                   &             &         &          &                                            \\
23 36 26.80 &  $+$33 02 15.1 & G128-084           &      K1\tablenotemark{a}  &                   &             &         &          &                                            \\                     
23 50 40.40 &  $+$17 20 40.5 & G030-024           &      K1                   &                   &             &         &          &                                            \\
\enddata
\label{tbl.speckle.observation}

\tablecomments{The site codes, K1, K2 and C1, indicate that targets
are observed at KPNO (2005.8625--2005.8692), KPNO
(2007.5876--2007.6076) and CTIO (2006.1882--2006.2001),
respectively. ``WDS'' in column 7 indicates the CPM binary separations
are from Washington Double Catalog \citep{Mason2001}. ``SB1'' and
``SB2'' in column 8 indicate a single-line spectroscopic binary
reported in \cite{Latham2002} and a double-line spectroscopic binary
reported in \cite{Goldberg2002}. ``SB'' indicates a spectroscopic
binary from \cite{Dawson2005}, but no orbital elements are
available. ``SB?'' for LHS 169 indicates a large $V_{r}$ variation
observed in \cite{Dawson2005}, and this system needs to be
re-observed. ``A'' and ``W'' in column 8 indicates targets have also
been observed by HST/ACS and HST/WFPC2 in \cite{Riaz2008} and
\cite{Gizis2000}.}

\tablenotetext{a}{Observed with wider filter (Johnson V: 545$\pm$85
                      nm) and lower microscope objective due to
                      character of target. Resolution limit for this
                      observation is estimated at
                      $\rho$~$<$~0\farcs05.}
\tablenotetext{b}{Known companion is too wide for detection here.}
\tablenotetext{c}{Known companion has too large a $\Delta$m for detection here.}

\end{deluxetable}


\clearpage
\begin{deluxetable}{llrllcrrccl}
\tablecaption{Resolved Subdwarf Binaries}
\tablewidth{17cm}
\tablehead{
\colhead{WDS Desig.}           &
\multicolumn{3}{c}{Discoverer} &
\colhead{Alternate}            &
\colhead{Epoch}                &
\colhead{$\theta$}             &
\colhead{$\rho$}               &
\colhead{$\Delta V$}            &
\colhead{Notes} \\
\colhead{$\alpha \delta$ (2000)} &
\multicolumn{3}{c}{Designation}  &
\colhead{Designation}            &
\colhead{2000.$+$}               &
\colhead{($\circ$)}              &
\colhead{(\arcsec)}              &
\colhead{(est.)}                 &
\colhead{} 
}
\startdata

03500$+$4325 & WSI &   68 & Ca,Cb & LHS0182   & 5.8685 & 172.5  & 0.62   & 0.5 & new   \\
04256$-$0651 & LDS &  842 &       & LHS0189   & 5.8685 & 280.6  & 2.81   & 2.1 &       \\
13317$-$0219 & HDS & 1895 &       & G062-044  & 6.1915 & 147.6  & 0.13   & 1.4 &       \\
19400$+$4256 & WSI &   69 &       & G125-026  & 7.6016 & 158.4  & 3.29   & 1.5 & new?  \\
\enddata

\label{tbl.new}
\end{deluxetable}


\begin{deluxetable}{lccc}
\tablecaption{Multiplicity Survey Coverage}
\tablewidth{13cm}
\tablehead{
\colhead{Samples}                 &
\multicolumn{3}{c}{Methods}         \\
\colhead{}                        &
\colhead{$V_{rad}$ Surveys}       &
\colhead{Speckle Interferometry}  &
\colhead{Blinking Plates}                 
}
\startdata
This work           &  14\%       &   100\% (at V band)         &  100\% \\
RG97                &  39\%       &   74\%  (at near-IR band)   &  100\% \\
\enddata
\label{tbl.our.RG97}
\end{deluxetable}


\begin{deluxetable}{ccc}
\tablecaption{Comparison of Multiplicity Results}
\tablewidth{13cm}
\tablehead{
\colhead{Stellar Type}     &
\colhead{A-G or G type}    &
\colhead{K \& M type}
}
\startdata
dwarfs     &   57\% (DM91)              &     37\% (RG97)      \\
subdwarfs  &   15\% to 30\% (Mixed)     &     26\% (This Work) \\
\enddata
\label{tbl:fraction}
\end{deluxetable}


\end{document}